\documentclass[prx,twocolumn,english,
floatfix,longbibliography]{revtex4-2}

\usepackage[utf8]{inputenc}
\usepackage[T1]{fontenc}
\usepackage{braket,color}
\usepackage{graphicx,enumerate,verbatim,bbold}
\usepackage{amsmath,amssymb,amsthm,float,mathrsfs}
\usepackage{dsfont}
\usepackage[normalem]{ulem}
\usepackage{lmodern}
\usepackage{hyperref}
\usepackage{braket}
\usepackage{multirow}
\usepackage{bbm}
\usepackage{physics}
\usepackage{bm}
\usepackage[normalem]{ulem}
\usepackage[caption=false]{subfig}
\usepackage{dcolumn}
\newcommand{\bqa}{\begin{eqnarray}}
\newcommand{\eqa}{\end{eqnarray}}

\newcommand{\be}{\begin{equation}}
\newcommand{\ee}{\end{equation}}

\newcommand{\flo}[1]{{\color[rgb]{0,0,0.6}{#1}}}
\newcommand{\luca}[1]{{\color[rgb]{0,.6,0}{#1}}}
\newcommand{\hil}[1]{{\color[rgb]{0.8,0,0.2}{#1}}}

\newcommand{\nguyen}[1]{{\color[rgb]{0.6,0,0.6}{#1}}}

\begin{document}
\title{
Robust implicit quantum control of interacting spin chains}

\author{Luca Stefanescu, Louis Edwards-Pratt, Jeremy O'Connor, Ezra Tsegaye, Nguyen H. Le, Florian Mintert}
\affiliation{Physics Department, Blackett Laboratory, Imperial College London, Prince Consort Road, SW7 2AZ, United Kingdom}

\begin{abstract}
Robust quantum control can achieve noise-resilience of quantum systems and quantum technological devices.
While the need for noise-resilience grows with the number of fluctuating quantities, and thus typically with the number of qubits, most numerically exact optimal control techniques are limited to systems of few interacting qubits.
This paper exploits quantum control that avoids explicit reference to quantum states in exponentially large Hilbert space. 
Exemplary control protocols for spin chains are discussed in terms of noise-resilient preparation of highly entangled states. 

\end{abstract} 
\maketitle


\section{Introduction}

Optimal control~\cite{ansel_introduction_2024} is a crucial technique in the development of quantum technological applications.
Since robust control \cite{weidner_robust_2024} can make the dynamics of quantum systems insensitive to fluctuations in system parameters,
it can be used to lower the requirements for calibration of a quantum device.
Given that the number of potentially fluctuating system parameters grows with increasing system size, the practical value of robust control increases with the ability to scale up quantum devices. 

Control pulses that achieve insensitivity of single-qubit dynamics to fluctuations in quantities like the resonance frequency are well established in nuclear magnetic resonance~\cite{levitt_composite_1986} and found applications in many other platforms including solid-state impurities~\cite{nobauer_smooth_2015}, superconducting qubits~\cite{carvalho_error-robust_2021,cykiert_robust_2024,Yi2024} and trapped ions~\cite{Timoney2008}.
Noise-resilience has also been achieved with high accuracy for two-qubit entangling gates~\cite{Kalra2014,Yi2024,Setiawan2023,Manovitz2017,webb_resilient_2018,shapira2018,Zarantonello2019,Manovitz2022,Barthel2023,Xie2023,Baum2021,Mitra2020,Mohan2023,Evered2023} and few-qubit gates~\cite{Manovitz2022,Kanaar2024,Liu2022},
but the development of control techniques to achieve noise-resilience gets rapidly more challenging with increasing system size~\cite{metz_self-correcting_2023}.

While robust control is often aimed at fluctuations of system parameters that would be finite in an ideal situation,
an aspect of practical importance in registers of interacting qubits are parasitic terms in a Hamiltonian, {\it i.e.} terms that ideally would vanish.
In superconducting circuits for example, robust control is a method of choice to mitigate the impact of such parasitic interactions~\cite{zhou_quantum_2023,watanabe2024}.

Numerous algorithms have been developed for optimal quantum control ranging from direct search \cite{caneva_chopped_2011}, gradient-based optimization  \cite{khaneja_optimal_2005,kosut_robust_2013,Leung2017,goerz_krotov_2019}, to machine learning inspired approaches \cite{bukov_reinforcement_2018,niu_universal_2019}. Robustness against noise in the model's parameters can be achieved in these approaches by optimizing an ensemble average over the noise distribution \cite{Propson2022}. These robust control techniques have been instrumental in designing error-resilient quantum gates and protocols \cite{Muller2022,koch_quantum_2022,allen_optimal_2017,le_scalable_2023}, but they are typically restricted to small to moderate system size.

A limiting factor in the design of control pulses that achieve noise-resilience in large quantum systems is the required effort to simulate the controlled dynamics, with typically exponential scaling in the number of qubits.
Notable exceptions include data-driven control~\cite{dive_situ_2018,sauvage_optimal_2020,mukherjee_bayesian_2020} with numerical simulations replaced by experimental data.
While such approaches can be used to achieve resilience even against uncharacterized noise,
it can be experimentally expensive, and the presence of drift in system parameters is a restrictive limiting factor~\cite{greenaway_variational_2024}. 
Counterdiabatic driving also allows for the construction of time-dependent control Hamiltonians, but the resultant Hamiltonians are often not practically realisable and need to be approximated dynamically~\cite{petiziol_fast_2018}. Techniques based on tensor networks exist for many-body optimal control \cite{Doria2011, jensen_achieving_2021,metz_self-correcting_2023}, but they are also approximate in nature and are subjected to Trotterization and truncation errors. 

The present work builds up upon a framework of implicit quantum control (IQC)~\cite{iqc2024} in which a quantum state $\ket{\Psi}$ is not specified explicitly, but implicitly in terms of an operator ${\cal I}$ to which it is an eigenstate.
Crucially, there are instances of Hamiltonians $H(t)$ and control problems for which this results in a sub-exponential, in particular quadratic, numerical effort to characterize the system dynamics.
In such instances, numerically exact optimal control \cite{iqc2024} is feasible despite the exponential effort required for the explicit specification of the time-evolving system state.

The goal of this paper is to demonstrate robust control within the framework of IQC
for spin chains with static noise in system parameters.
Given the fundamental difficulty to overcome the general exponential scaling of a quantum system's Hilbert space with the number of qubits, the approach~\cite{iqc2024} used here has a restricted range of applicability that is easily broken by parasitic Hamiltonian terms.
We thus develop a perturbative treatment that overcomes this limitation, and we provide explicit control solutions for spin chains with parasitic interactions of up to 20 spins.

We demonstrate the effectiveness of this approach by designing robust control solutions to realize a resource state for measurement-based quantum computation 
and to implement a Heisenberg-limited sensing protocol.
Notably, our findings reveal that $99.9\%$ fidelity can be maintained even in the presence of significant errors—up to 5\%—in all the spin-spin coupling strengths.

\section{Background}\label{sec:bac}

\subsection{Robust quantum control}

The goal of optimal quantum control is the design of a time-dependent Hamiltonian, such that the propagator induced by this Hamiltonian approximates desired properties as well as possible.
Commonly pursued tasks include the realization of a targeted state starting from some initial state or the implementation of a quantum gate.

The design of a Hamiltonian with an optimal time-dependence is typically an iterative process involving the evaluation of an objective function for a given time-dependence and the construction of a modified time-dependence with the prospect of yielding a higher value of the objective function than previously tested time-dependencies.

In order to achieve noise-resilience, one can consider an ensemble of Hamiltonians, such that a given system parameter adopts a different value for each ensemble member. The goal of the optimization is then to find one single time-dependence that applies to every ensemble member, such that the ensemble averaged objective function is maximized.
Given a sufficiently large ensemble, the controlled system is robust against fluctuations of the considered parameter for parameter 
values within the interval chosen for the ensemble.

\subsection{
Implicit quantum control}
\label{sec:invcont}

Given the generally exponential scaling of the Hilbert space with the number of qubits,
it can be advantageous to characterize quantum states $\ket{\Psi(t)}$
in terms of an eigenvalue relation
\be
{\cal I}(t)\ket{\Psi(t)}=\gamma\ket{\Psi(t)}\ ,
\ee
formulated in terms of a unitarily time-evolving Hermitian operator ${\cal I}(t)$~\cite{iqc2024}.

Given a time-dependence, following the von Neumann equation
\be
\dot{\cal I}(t)=-i[H(t),{\cal I}(t)]\ ,
\label{eq:eqom}
\ee
any non-degenerate eigenstate of ${\cal I}$ provides a solution to the time-dependent Schr\"odinger equation~\cite{lewis_exact_1969}.
With an initial condition ${\cal I}(0)$ having a non-degenerate eigenstate $\ket{\Psi(0)}$ and a control target ${\cal I}_T$ having an eigenstate $\ket{\Psi_T}$ to the same eigenvalue,
an optimal control process with an objective function that is maximized exactly for the target ${\cal I}_T$, such as the infidelity
\be
{\cal J}({\cal I},{\cal I}_T)=1-\frac{\mbox{Tr}({\cal I}{\cal I}_T)}{\mbox{Tr}({\cal I}_T^2)}\ ,
\label{eq:infid}
\ee
yields a time-dependent Hamiltonian that achieves the state transfer from the initial state $\ket{\Psi(0)}$ to the target state $\ket{\Psi_T}$ \cite{iqc2024}.

In order to solve the von Neumann equation numerically, it is helpful to expand the Hamiltonian and the operator ${\cal I}(t)$ in terms of two sets of time-independent operators $\{\mathfrak{h}_j\}$ and $\{\mathfrak{a}_j\}$.
With the expansions
$H(t)=\sum_jh_j(t)\mathfrak{h}_j$ and
${\cal I}(t)=\sum_ja_j(t)\mathfrak{a}_j$ with time-dependent coefficients $h_j(t)$ and $a_j(t)$,
and the closed commutator relations
\be
[\mathfrak{h}_j,\mathfrak{a}_k]=\sum_{l=1}^d\lambda_{kl}^j\mathfrak{a}_l\ ,
\label{eq:closed}
\ee
the von Neumann equation yields the differential equation
\be
\dot a_i(t)=\sum_jK_{ij}(t)a_j(t)
\hspace{.3cm}
\mbox{with}
\hspace{.3cm}
K_{ij}(t)=\sum_lh_l(t)\lambda_{ji}^l\ .
\label{eq:deq}
\ee

The dimension of the dynamical problem to be solved is thus given by the number of terms $\mathfrak{a}_j$ in the expansion of ${\cal I}(t)$ that is required to ensure that the commutator relation Eq.\eqref{eq:closed} is closed.
While generally, this can also be exponential in the number of qubits, and even quadratic in the dimension of the Hilbert space,
there are classes of Hamiltonians and associated operators ${\cal I}(t)$ that show polynomial and notably quadratic scaling with system size~\cite{iqc2024}.

This approach can be placed under the umbrella of shortcuts to adiabaticity \cite{guery-odelin_shortcuts_2019}, with a state vector $\ket{\Psi(t)}$ following the dynamics of the operator ${\cal I}(t)$, but no requirement to follow the dynamics of $H(t)$ adiabatically.
As such, it allows the preparation of ground states of interacting models that are challenging to realize, such as those with long-range interactions or three-or-more body interactions, using a control Hamiltonian that requires only nearest-neighbor two-body interactions.

\section{Linear spin chain}
\label{sec:spinchain}

The system Hamiltonian
\be
H(t)=\sum_{j=1}^{n-1}g_jX_jX_{j+1} +\sum_{j=1}^{n}f_j(t)Z_j+ \hspace{-0.1cm}
\sum_{j\in\{1,n\}}w_j(t)X_j
\label{eq:ham}
\ee
for a linear spin chain with nearest-neighbor interactions in terms of the Pauli $X$ operators, single-spin energies in terms of the Pauli $Z$ operators and the Pauli $X$ operators on the end-spins of the chain is characterized by the set of operators $\{\mathfrak{h}_j\}$ with the terms $Z_j$, $j\in[1,n]$, $X_jX_{j+1}$, $j\in[1,n-1]$, $X_1$ and $X_n$.
A good choice for efficiently evaluable time-evolving operator ${\cal I}(t)$ is given by the set of operators $\{\mathfrak{a}_j\}$ that contains all terms of the set $\{\mathfrak{h}_j\}$, and all nested commutators with terms in $\{\mathfrak{h}_j\}$ under the constraint that the set $\{\mathfrak{a}_j\}$ is linearly independent.
Since this set has only $2n^2+3n+1$ elements~\cite{iqc2024}, Eq.~\eqref{eq:eqom} can be represented in the form of Eq.~\eqref{eq:deq} in a vector space of quadratic size. The explicit expression for all the operators in the set $\{\mathfrak{a}_j\}$ is given in the Appendix.

The system parameters $f_j(t)$ represent the single-qubit resonance frequencies, that may have a time-dependence, {\it e.g.} due to a time-dependent magnetic field.
The parameters $w_1(t)$ and $w_n(t)$ represent driving on the first and last spin of the chain and the parameters $g_j$ represent the coupling strength between two neighboring spins.
In a practical implementation the system parameters 
may be subject to fluctuations.
While single-qubit parameters are typically characterized with high precision, the coupling strengths between qubits often have significant uncertainties—especially in platforms with complex coupling schemes. In superconducting devices, for example, two-qubit errors tend to be ten times higher than single-qubit errors \cite{krinner_demonstration_2020,arute_quantum_2019}.
We thus consider the noise model
\begin{equation}\label{eq:noise}
    g_j=g+\epsilon_j,
\end{equation}
for the individual interaction strengths, 
where $\epsilon_j$ represents the fluctuation specific to each interaction and is assumed to be distributed uniformly at random in a small error interval $[-\Delta \epsilon, \Delta \epsilon]$.



Given an ensemble of Hamiltonians $H_i(t)$ of the form of Eq.~\eqref{eq:ham}, with each ensemble member adopting different values of the systems parameters,
Eq.~\eqref{eq:eqom} defines an ensemble of time-evolving operators ${\cal I}_i(t)$.
Each of these has the same initial condition ${\cal I}_i(0)={\cal I}(0)$, but given the different values of the system parameters, the final values ${\cal I}_i(T)$ will generally not be the same.
The deviation of these final values from the control target ${\cal I}_T$ can be characterized in terms of an average infidelity
\be\label{eq:avginfid}
\bar{\cal J}=\frac{1}{M}\sum_{i=1}^{M}{\cal J}({\cal I}_i,{\cal I}_T)\ ,
\ee
with the regular infidelity defined in Eq.~\eqref{eq:infid} and $M$ the number of members in the ensemble. 

\section{Robust implicit quantum control}
\label{sec:rob}
With an ensemble of Hamiltonians of the form Eq.~\eqref{eq:ham}, the approach of IQC \cite{iqc2024} sketched in Sec.~\ref{sec:invcont} allows for an exact treatment of parameter fluctuations without resort to perturbative approximations. 

With $n-1$ interaction constants subject to fluctuations, the ensemble average of the objective function (Eq.\eqref{eq:avginfid}) implies an average in an $n-1$-dimensional space.
With any uniform distribution of sampling points, this implies an exponential and thus prohibitively expensive effort.
A more efficient method is based on sampling over randomly chosen points,
similarly to stochastic gradient descent in big data machine learning~\cite{sra_optimization_2012}.

%
%

\subsection{Cluster state preparation}

Preparation of the ground state of the operator
\be\label{eq:hc}
H_C= Z_1X_2 + \sum_{j=1}^{n-2}X_jZ_{j+1}X_{j+2} + X_{n-1}Z_n
\ee
is of interest because it is the cluster state, a resource state for measurement-based quantum computation \cite{briegel_measurement-based_2009,walther_experimental_2005}.
The initial condition ${\cal I}(0)=\sum_iZ_i$ and the control target ${\cal I}_T=H_C$ are suitable choices for the task of realizing the ground state of $H_C$ since the initial state of ${\cal I}(0)$ is separable and thus typically preparable in practice, and the target ${\cal I}_T$ can be reached with this initial condition and the dynamics induced by the physical Hamiltonian  $H(t)$ (Eq.~\eqref{eq:ham}) \cite{iqc2024}.

To achieve robust control, the average infidelity in Eq.~\eqref{eq:avginfid} is optimized using the  average gradient from a sample of 60 points selected uniformly at random within the error range. After completing the optimization, the robustness of the optimized pulse is validated 
with an averaged infidelity  over  $10^3$ new random points.

\begin{figure}
    \centering
    \includegraphics[width=0.9\linewidth]{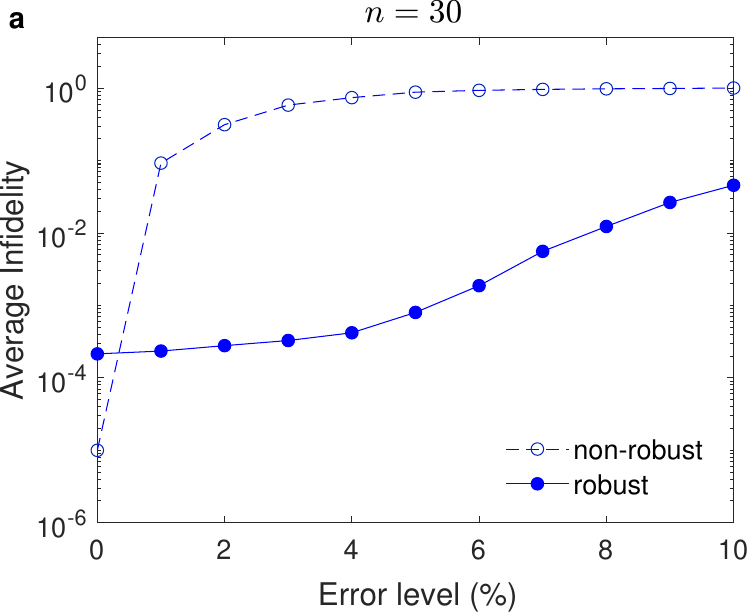}
    \includegraphics[width=0.9\linewidth]{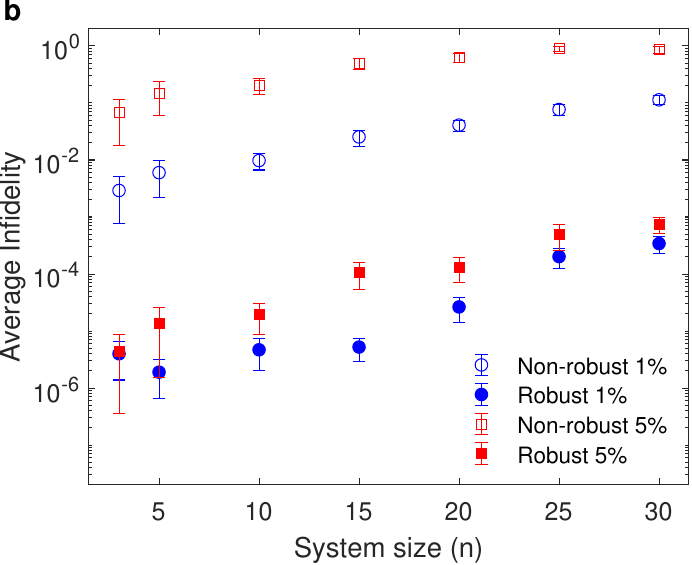}    
    \caption{(a) Average infidelities for preparing a 30-qubit cluster state, obtained with  pulses optimized without  robustness and with robustness, versus increasing error levels $\Delta \epsilon/g$. (b) Average infidelity, with error bars indicating the standard deviation of the infidelity distribution, for preparing the cluster state with robustness (solid) and without robustness (empty) for 1\%  (square) and 5\% (circle) error level in the coupling strength.}
    \label{fig:infid}
\end{figure}

Fig.~\ref{fig:infid}a depicts the infidelity for a chain with $n=30$ spins as a function of the error level $\Delta \epsilon/g$.
The empty circles correspond to a control solution optimized for the noiseless case.
It achieves the targeted infidelity of $10^{-5}$ for vanishing noise, but the results deteriorate rapidly with increasing noise level.
The filled circles correspond to a control solution optimized for noise of up to $5\%$ in the coupling constants.
Achieving noise resilience comes at the price of a slightly increased infidelity for the noiseless case, but, in turn, infidelities lower than $10^{-3}$ are obtained in the entire $5\%$ interval.


Fig.~\ref{fig:infid}b depicts average infidelity as a function of system size $n$ with robustness for fluctuations in the coupling constants or $1\%$ (blue) and $5\%$ (red).
The open circles correspond to pulses optimized for the noiseless case, and the error bars depict the width of the distribution of infidelities contributing to the average.
While it is possible to obtain infidelities on the order of $10^{-2}$ for systems of 3 spins with pulses optimized for the noiseless case, their performance deteriorates rapidly with increasing system size, highlighting the need for engineered noise-resilience in larger systems.
Control solutions designed to be noise-resilient, on the other hand, result in infidelities that grow very slowly with system size and even chains with $n=30$ spins feature infidelities as low as $10^{-3}$.

\begin{figure}
    \centering
    \includegraphics[width=\linewidth]{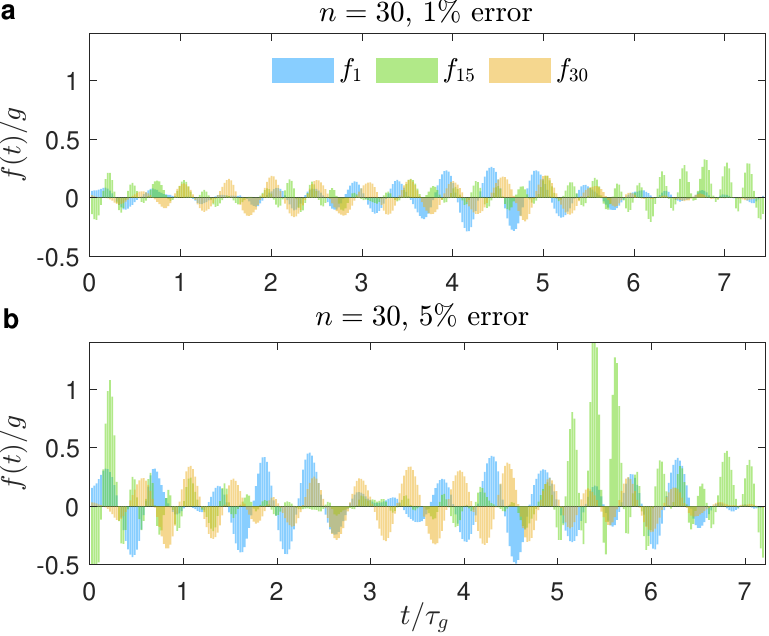}
    \caption{Pulse shapes of $f_1,f_{15}$, and $f_{30}$, for the cluster state preparation in a chain with $n=30$ spins, optimized for $1\%$ error level (a) and $5\%$ error level (b), plotted as functions of time in multiples of the interaction time scale $\tau_g$. The amplitude of the pulses increases with the error level as expected due to the increasing difficulty of correcting the effect of larger fluctuation.}
    \label{fig:cluster_pulses}
\end{figure}

The extent to which noise-resilience can be achieved depends on the duration of controlled dynamics and on properties such as the spectral width of the control functions $f_j(t)$ and $w_j(t)$.
Achieving infidelities as low as in Fig.~\ref{fig:infid}b requires a duration that scales linearly with the system size, $T\sim n \tau_g/4$ where $\tau_g=2\pi/g$ is the characteristic timescale associated with the interaction strength.
Given that an optimization is initialized with randomly chosen time-dependencies, the spectral width of the optimized pulses can be unnecessarily broad.
The spectral width can be reduced by a process of smoothing and subsequent use of the smooth pulses as initialization of a new optimization.
Exemplary time-dependencies for the control functions $f_{1}, f_{15}$ and $f_{30}$ in a chain with $n=30$ spins are depicted in Fig.~\ref{fig:cluster_pulses}a for 1\% error level and in Fig.~\ref{fig:cluster_pulses}b for 5\% error level.
The optimizations do not include a penalty for high amplitudes, but since the evolution towards the targeted entangled state is limited by the strength of the interactions,
there is no reason for the optimization to increase amplitudes beyond what is necessary.
The maximum amplitudes obtained are thus on the order of $g$ and they grow only moderately with increasing noise level.
All these features exemplified in Fig.~\ref{fig:cluster_pulses} hold similarly for all the driving functions $f_j(t)$ and $w_j(t)$ as provided in Ref.~\cite{zenodo}.

\subsection{Quantum sensing}\label{sec:sensing}



Entangled states such as the GHZ state can be utilized for ultra-sensitive quantum measurement reaching the Heisenberg precision limit \cite{degen_quantum_2017}. A quantum sensing experiment typically requires three crucial steps:
preparation of an entangled state, free phase evolution and measurement of the phase shift.
In practice the last step requires a manipulation of the system state, such that the phase shift can be measured in terms of accessible single-qubit measurements.
In addition to noise-resilience of the state preparation, a noise-resilient sensing protocol thus also requires a noise-resilient implementation of the final measurement.


Given that most sensing protocols are implemented with superpositions of $Z$ eigenstates, the discussion in this section is based on a Hamiltonian of the form of Eq.~\eqref{eq:ham} with a $\pi/2$ $Y$ rotation, ({\it i.e.} with $X$ replaced by $-Z$, and $Z$ replaced by $X$).
The GHZ state
\be
\ket{\Psi_G}=\frac{1}{\sqrt{2}}\left(\ket{0}^{\otimes n}+\ket{1}^{\otimes n}\right),
\ee
with the single-qubit Z eigenstates $\ket{0}$ and $\ket{1}$, is the ground state of the operator
\be
H_G=-\sum_{j=1}^{n-1}Z_j Z_{j+1}-\prod_{j=1}^n X_j\ ,
\ee
so that $\mathcal{I}_T=H_G$ together with the initial condition $\mathcal{I}(0)=\sum_j X_j$ is a suitable control target.


The two states $\ket{0}^{\otimes n}$ and $\ket{1}^{\otimes n}$ are both eigenstates of the system interaction $\sum_jg_jZ_jZ_{j+1}$ to the eigenvalue $\sum_jg_j$.
The interaction does thus not result in a relative phase in the GHZ state so that there is no need to engineer noise-resilience for the step of the free dynamics in the sensing protocol.

Any phase shift $\theta$ of a single qubit due to an additional field will contribute to the overall phase shift in the state
\be
\ket{\Psi_G(\theta)}=\frac{1}{\sqrt{2}}\left(\ket{0}^{\otimes n}+e^{in\theta}\ket{1}^{\otimes n}\right)\ .
\ee

This phase shift can be measured in terms of the observable $O=\prod_{j=1}^n X_j$,
and this n-qubit observable can be mapped onto a single-qubit observable with a second interval of controlled dynamics.
 
The remaining goal is thus to find a system Hamiltonian with controlled time-dependence that results in a propagator $U$ such that $UOU^\dagger=X_1$.
This goal can be achieved with the control problem defined in terms of the initial condition ${\cal I}(0)=O$ and the control target ${\cal I}_T=X_1$.

\begin{figure}
    \centering
\includegraphics[width=0.9\columnwidth]{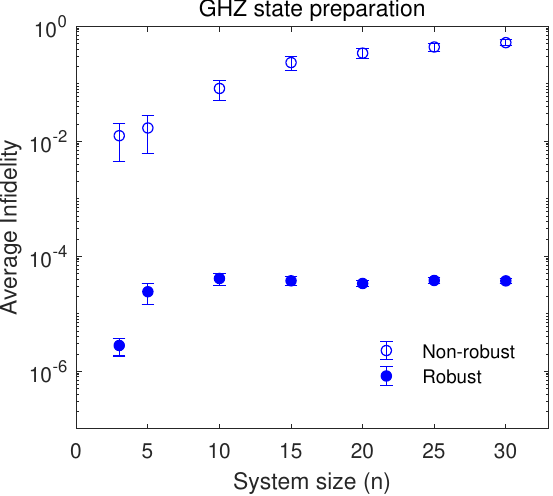}
    \caption{ Average infidelity for GHZ state preparation at 5\% error level, obtained with pulses optimised without robustness (empty circle) and with robustness (solid circle). The error bar indicates the standard deviation of the infidelity distribution.}
    \label{fig:ghz}
\end{figure}

\begin{figure}
    \centering
\includegraphics[width=0.9\columnwidth]{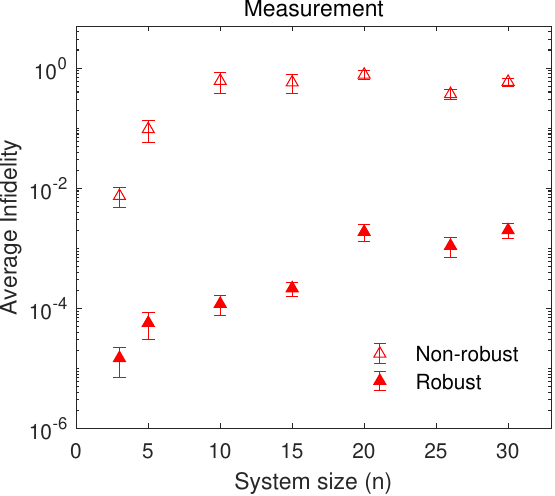}
    \caption{
    Average infidelity for the mapping from the collective operator $O\equiv \prod_j X_j$ to the single qubit operator $X_1$ at 5\% error level for the pulses optimised without robustness (empty triangle) and with robustness (solid triangle).}
    \label{fig:measure}
\end{figure}

Figure.~\ref{fig:ghz} shows the average infidelity achieved for preparing the GHZ state with and without robustness at a 5\% error level. As with the cluster state, the impact of fluctuations is pronounced for the pulses optimized without robustness. However, with robust control, an improvement of four orders of magnitude in average infidelity is achieved. The duration of the optimized pulses is $T\sim n \tau_g/4$, which is consistent with the time required to generate the long range correlation in the GHZ state with local interaction. Figure \ref{fig:measure} shows the average infidelity achieved for the transformation required for the measurement step, evaluated with pulses optimized with and without robustness  at a 5\% error level. Robust control again delivers a drastic improvement, reducing average infidelity by at least three orders of magnitude.
Here, a duration $T\sim n \tau_g/2$ seems to be required to reach 99.9\% fidelity, which is a factor of two longer than for the state preparation.

\section{Robustness against fluctuation outside the closed set}\label{sec:perturbative}

In addition to control terms in the Hamiltonian, there might also be parasitic  terms due to experimental imperfection. 
A well-known example of this in superconducting platforms is the parasitic $ZZ$ interaction \cite{ni_scalable_2022} as an extra term to the system Hamiltonian (Eq.\eqref{eq:ham}).
As long as the commutator relation in Eq.~\eqref{eq:closed} is satisfied for all the terms $\mathfrak{h}_j$ including the parasitic terms in the system Hamiltonian, the present approach can be used to design robust control as exemplified above.
Generically, however, Eq.~\eqref{eq:closed} is not satisfied for all parasitic terms in the Hamiltonian, and an extension of the set of operators $\{\mathfrak{a}_j\}$ results in unfavourable scaling. Adding the parasitic terms $Z_jZ_{j+1}, 1\leq j \leq n$ to the control Hamiltonian Eq.~\eqref{eq:eqom}, for example, leads to an exponentially large set.

Since parasitic terms are expected to be weak as compared to the desired terms in the Hamiltonian, it is a viable perspective to take them into account perturbatively.
To this end, it is helpful to express the full Hamiltonian as $H(t) = H_0(t) + H_P$ 
where $H_0(t)$ is comprised only of terms that satisfy the commutator relation Eq.~\eqref{eq:closed} with a sufficiently small set $\{\mathfrak{a}_j\}$, and $H_P$ contains the parasitic terms to be treated perturbatively.

With an operator ${\cal I}(t)$ satisfying Eq.~\eqref{eq:eqom} and the propagator $U_0(t,0)$ induced by $H_0(t)$, the operator
\be
\label{eq:perttrans}
{\cal I}_1(t) = U_0^\dagger(t,0){\cal I}(t)U_0(t,0)
\ee
satisfies the equation of motion
\be
\dot{\cal I}_1(t)=-i[
\tilde{H}_P(t),{\cal I}_1(t)]\ ,
\label{eq:eqomI1}
\ee
with
$\tilde{H}_P(t) = U_0^\dagger(t,0)H_PU_0(t,0)$.
This can be taken as starting point for a perturbative solution in terms of a
Dyson series,
but an explicit construction of $\tilde{H}_P(t)$ is required for that. 

Since the situations of interest are such that the propagator $U_0(t,0)$ is a too high-dimensional object to be constructed explicitly, it is essential that the operator $\tilde H_P(t)$ is not constructed via matrix multiplication, but that it is constructed via integration of an equation of motion.
The equation of motion that $\tilde H_P(t)$ satisfies reads
\be
\dot{\tilde H}_P(t)=i[U_0^\dagger(t,0)H_0(t)U_0(t,0),\tilde H_P(t)]\ .
\ee
Explicit construction of the operator $U_0^\dagger(t,0)H_0(t)U_0(t,0)$ seems infeasible, so that direct integration of this equation of motion is not a good way forward.

The operator
\be
Q(\tau)=U_0^\dagger(t,t-\tau)H_PU_0(t,t-\tau)\ ,
\label{eq:eqomQ}
\ee
however, has an equation of motion that can be integrated in practice, and it reads
\be
\frac{\partial Q(\tau)}{\partial\tau}=i[H_0(t-\tau),Q(\tau)]\ .
\label{eq:eqomQ}
\ee
Since the equality $\tilde H_P(t)=Q(t)$ holds, one can thus integrate Eq.~\eqref{eq:eqomQ} for any value of $t$ with $\tau$ starting at $\tau=0$ and ending at $\tau=t$ to obtain $\tilde H_P(t)$.

The effort required for the integration of Eq.~\eqref{eq:eqomQ} depends on the initial condition $Q(0)=H_P$.
Integration is certainly efficient if $H_P$ can be expanded into a set of operators with the closed commutator relationship in Eq.~\eqref{eq:closed}, such as the set $\{\mathfrak{a}_k\}$.
This condition, however, is only sufficient and not necessary.
Since any solution $R(\tau)$ of Eq.~\eqref{eq:eqomQ} is of the form $R(\tau)=U_0^\dagger(t,t-\tau)R(0)U_0(t,t-\tau)$,  the product of several solutions $R_j(\tau)=U_0^\dagger(t,t-\tau)R_j(0)U_0(t,t-\tau)$ is also a solution of Eq.~\eqref{eq:eqomQ}. It can thus be practically possible to construct $\tilde H_P(t)$, even if $H_P$ cannot be expanded into a small set of operators that satisfy Eq.~\eqref{eq:closed}.

Given an explicit construction of $\tilde H_P(t)$, the perturbative solution to Eq.~\eqref{eq:eqomI1} reads
\be
\label{eq:firstcorrection}
{\cal I}_1(t) = {\cal I}_1(0) -i\int_{0}^td\tau[\tilde H_P(\tau),{\cal I}_1(0)],
\ee
with ${\cal I}_1(0)={\cal I}(0)$,
and this perturbative impact of the parasitic terms in $H_P$ is the basis of the subsequent analysis.

\subsection{Perturbative correction}
With $H_P$ expanded in a  set of operators
\be
H_P=\sum_j \lambda_j \mathfrak{k}_j(0),
\ee
where the strength parameter $\lambda_j$ might be unknown as in the cases of parasitic interaction, the time evolved perturbing Hamiltonian can be written as
\be
\tilde H_P(t)=\sum_j\lambda_j \mathfrak{k}_j(t),
\ee 
where
\be
\mathfrak{k}_j(t)=U_0^\dagger(t,0)\mathfrak{k}_j(0)U_0(t,0)\ . 
\label{eq:dynpert}
\ee
Provided that the time-dependent operators $\mathfrak{k}_j(t)$ can be constructed, they can be expanded in some basis set of operators $\{\mathfrak{K}_p\}$,
\be
\mathfrak{k}_j(t)=\sum_pv_{jp}(t)\mathfrak{K}_p\ .
\label{eq:expand}
\ee
The explicit time-dependence of $\tilde H_P(t)$ then reads
\bqa\label{eq:h1lp}
\tilde H_P(t)=\sum_{jp}\lambda_j v_{jp}(t)\mathfrak{K}_p\ .
\eqa
With $\mathcal{I}(0)$ expanded in the small set as $\mathcal{I}(0)=\sum_l a_l \mathfrak{a}_l$, the operator-valued part of the integral $\int dt\ [\tilde H_P(t),{\cal I}(0)]$ that determines the impact of parasitic Hamiltonian terms is given by the commutators $[\mathfrak{K}_p,\mathfrak{a}_l]$.
Since the set of all these commutators is not necessarily linearly independent, it is helpful to expand them as
\be\label{eq:lab}
[\mathfrak{K}_p,\mathfrak{a}_l]=i\sum_q\eta_{lq}^p\mathfrak{b}_q,
\ee
where $\mathfrak{b}_q$ are linearly independent operators.

The first order impact of the parasitic Hamiltonian terms at time $T$ then reads
\be
\int_0^T dt\ [\tilde H_P(t),{\cal I}(0)]=\sum_{jplq}a_l\eta_{lq}^p\mathfrak{b}_q\lambda_j \int_0^{T} dt\  v_{jp}(t)\ .
\ee
This impact vanishes for all $\lambda_j$ if each of the terms
\be
\sum_{lp}\eta_{lq}^p\int_0^T dt\, v_{jp}(t)
\ee
vanish. This is guaranteed if the sum
\be\label{eq:constraint}
{\cal C} = \sum_{jq}\left|\sum_{lp}\eta_{lq}^p\int_0^T dt\, v_{jp}(t)\right|^2
\ee
vanishes. The effect of the perturbation is then eliminated to first order for any value of $\lambda_j$ as long as they are sufficiently small for the perturbative treatment to remain valid. This is crucial for robustness of the dynamics against shifts in $\lambda_j$.

A time-dependent Hamiltonian such that the control target is reached in the absence of parasitic Hamiltonian terms, and such that ${\cal C}$ vanishes, thus ensures that the control target is reached in a noise-resilient fashion, {\it i.e.} there is no first-order impact of parasitic Hamiltonian terms to the system dynamics.
Hamiltonians with such a time-dependence can be found with an optimization using an objective function that includes a contribution depending on ${\cal C}$ or with an objective function evaluated in the absence of parasitic Hamiltonian terms and an additional constraint that ${\cal C}$ is smaller than some tolerable maximum values.

\subsection{Parasitic ZZ Interaction}\label{sec:ZZ}

 The perturbative correction described above is now utilized for mitigating the effects of the parasitic $ZZ$ interaction, which is commonly encountered in superconducting qubits. The unperturbed control Hamiltonian, $H_0(t)$ is given by Eq.~\eqref{eq:ham}, while the parasitic term is 
\be
H_P=\sum_{j=1}^{n-1} \lambda_j Z_j Z_{j+1},
\ee
where $\lambda_j$ is the small, but unknown, strength of the parasitic interaction. Our goal is to find an optimal control solution for realizing the cluster state that remains robust against this parasitic interaction. The initial condition is set as $\mathcal{I}(0)=\sum_j Z_j$ and the control target is $\mathcal{I}_T=H_C$ (Eq.~\eqref{eq:hc}).

The set \{$Z_j Z_{j+1}$\} is not part of the quadratic set $\{\mathfrak{a}_j\}$, which satisfies the closed commutator relations in Eq.~\eqref{eq:closed}. Including \{$Z_j Z_{j+1}$\} in \{$\mathfrak{h}_j$\} leads to an exponentially large set. The set \{$Z_j$\}, however, is in the small set $\{\mathfrak{a}_j\}$, allowing \{$Z_jZ_{j+1}$\} to be propagated as a product of efficiently propagated operators. Specifically, $U_0^\dagger(t,0)Z_jU_0(t,0)$ can be obtained efficiently by integrating the differential equation Eq.~\eqref{eq:deq} with the initial condition $Q(0)=Z_j$. 
This can be expressed as $\sum_k z_{jk}(t) \mathfrak{a}_k$ where $z_{jk}(t)$ are the expansion coefficients. The time-evolved parasitic Hamiltonian is then 
\be
\tilde H_P(t)= \sum_{jkl}\lambda_j z_{jk}(t)z_{j+1,l}(t)\mathfrak{a}_k\mathfrak{a}_l,
\ee
which has the same structure as the expansion in Eq.~\eqref{eq:h1lp} with $\{\mathfrak{K}_p\}=\{\mathfrak{a}_j\mathfrak{a}_k\}$. The linearly independent set $\{\mathfrak{b}_q\}$ for expanding the commutator relation in Eq.~\eqref{eq:lab} is then obtained, from which the constraint $\mathcal{C}$ in Eq.~\eqref{eq:constraint} can be evaluated.
For achieving high fidelity state preparation which is robust against the parasitic interaction, the infidelity $\mathcal{J}$ from Eq.~\eqref{eq:infid}, calculated for the dynamics of the unperturbed Hamiltonian $H_0(t)$ in Eq.~\eqref{eq:ham}, and the constraint $\mathcal{C}$ must be minimized together. This is done by minimizing the objective function $\mathcal{J}+\mathrm{w}\,\mathcal{C}$ where  $\mathrm{w}$ is a properly chosen weight (see Appendix).

The robustness of the optimal pulses obtained with perturbative correction can be verified for small system sizes by calculating the average infidelity within the error hypercube using the exact Lanczos propagator. In this calculation, $H_P$ is included in the control Hamiltonian, and the uncertain parameters $\lambda_j$ are allowed to vary within the range $[-\Delta \lambda, \Delta \lambda]$ where $\Delta \lambda$ is the error level. The von-Neumann equation is propagated exactly, and the infidelity distribution is obtained from a sample of $10^3$ random points within the hypercube (see Sec.~\ref{sec:rob}). This exact computation is done in the full, exponentially large, Hilbert space, and is thus limited to system size of $n\leq 10$. For larger system size, robustness can be inferred from the magnitude of the constraint $\mathcal{C}$ as it is a measure of the perturbation's effect.



\begin{figure}
    \centering
\includegraphics[width=1\columnwidth]{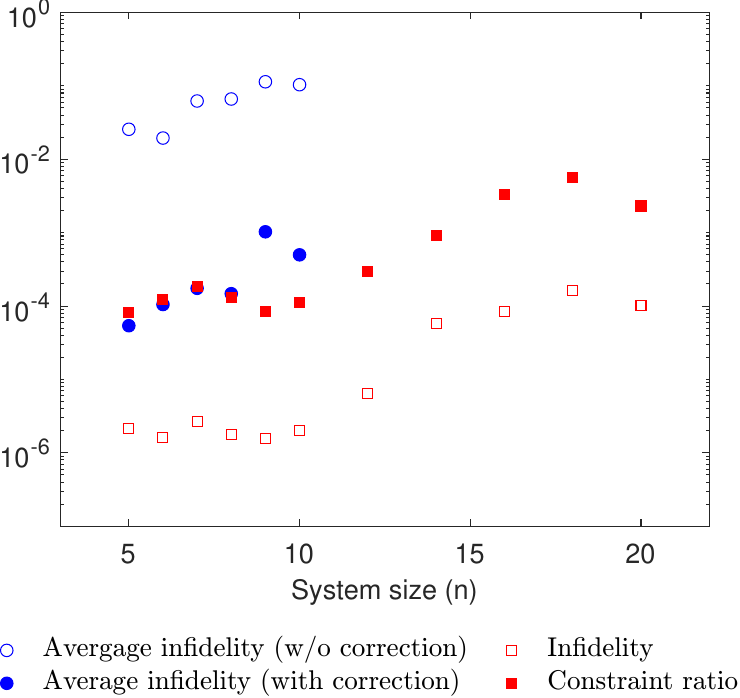}
\caption{Average infidelity of the cluster state preparation, at the error level $\Delta \lambda/g=5\%$, for time dependence  optimized without perturbative correction (empty blue circle) and with perturbative correction (solid blue circle). These are calculated with exact propagation in the exponentially large Hilbert space for $n\leq 10$.
Also shown for larger system size is the infidelity obtained for the unperturbed dynamics with the time dependence optimized with perturbative correction (empty red square) and the  ratio $\mathcal{C}_1/\mathcal{C}_0$ (solid red squares), where $\mathcal{C}_1$ is the constraint obtained for the time dependence optimized with perturbative correction and $\mathcal{C}_0$ the constraint obtained for the time dependence optimized without perturbative correction.}
    \label{fig:ZZ}
\end{figure}

Figure.~\ref{fig:ZZ} illustrates the average infidelity for the cluster state preparation at the error level $\Delta \lambda/g=5\%$ where $g$ is the interaction strength. 
The blue solid circles show infidelities for time-dependent Hamiltonians optimized to achieve resilience against parasitic $ZZ$ interactions,
whereas the blue empty circles correspond to Hamiltonians that are optimized for the state preparation only in the absence of parasitic interactions.
The reduction of infidelity by about two orders of magnitude underpins the resilience that can be achieved with suitably optimized time-dependence.
Given the exponential growth of the Hilbert space, it becomes prohibitively expensive to evaluate exact state infidelities for systems with more than $10$ spins.
The infidelities (Eq.~\eqref{eq:infid}) in the absence of parasitic interactions and the constraint ${\cal C}$, however, can be evaluated also for larger systems. This infidelity is shown by the empty red squares in Fig.~\ref{fig:ZZ}. To illustrate the suppression of  parasitic interaction achieved with perturbative correction, the ratio $\mathcal{C}_1/\mathcal{C}_0$ between the constraint $\mathcal{C}_1$, obtained for the time dependence optimized with perturbative correction, and the constraint  $\mathcal{C}_0$, obtained for the time dependence optimized without perturbative correction, is shown by the solid red squares. Despite  the rapid growths of the Hilbert space with $n$ and the the increasing number of parasitic interactions, there is only a moderate increase in both the infidelity and the constraint ratio. This indicates that state preparation can be achieved with high accuracy and robustness also in the longer spin chains.


\section{Conclusions \& Outlook}

The necessity to devise an accurate model is a decisive limiting factor in open-loop optimal control methods for quantum systems with more than a few qubits.
While noise-resilience is of practical importance for few-qubit systems, it thus appears to be even of fundamental importance for larger qubit registers.

While it seems plausible that an average over an ensemble of moderate size can be used for robust control of small systems with a small number of fluctuating parameters, the present work provides evidence that systems with a substantial number of fluctuating parameters also do not require excessively large ensembles in order to achieve a degree of noise-resilience that is helpful in practice.

The examples of state preparation and measurement discussed here can only exemplify the use of robust IQC, but they can not describe its use exhaustively.
In particular, syndrome measurements in quantum error correction are typically complicated many-body observables that need to be mapped onto single qubit observables in practice.
Since any imperfection in such operations require further layers of quantum error correction, any noise-resilient implementation of such mappings is highly desirable.

While the original goal of IQC is control of systems comprised of many qubits, it can also be beneficial for control of small systems interacting with an environment.
In this case, it is not essential to reduce the description of the system dynamics as it is in the case of a system with an exponentially large Hilbert space.
The interaction with an environment, however, requires an efficient treatment, and common techniques such as Markovian master equations have limited use for the design of optimal control protocols.
The present framework, in turn, can help to include the most relevant environmental degrees of freedom with tolerable computational overhead,
and to design optimal control protocols that achieve the desired goal while also decoupling the system from detrimental environmental influence.

Given the experimental advances in coherent control over systems of increasing size, and the correspondingly growing demand for suitable control protocols, the ability to devise robust control for systems beyond the two-qubit paradigm can thus provide valuable support for a broad range of experiments and quantum technological platforms.

\subsection*{Acknowledgements}
This work was supported by the U.K. Engineering and Physical Sciences Research Council via the EPSRC Hub in Quantum Computing and Simulation (EP/T001062/1) and the UK Innovate UK (project number 10004857).
We are indebted to stimulating discussions with Modesto Orozco-Ruiz. The optimised pulse and computer code for this work are available without restriction \cite{zenodo}.

\appendix

\section{Subalgebra of the problem}

The subalgebra is the set of operators generated by nested commutators of $\{\mathfrak{h}_j\}$ with $\{\mathfrak{a}_j\}$. If we start with $\{\mathfrak{a}_j\} = \{Z_j\}$ which is enough to express ${\cal I}(0)$ and $\{\mathfrak{h}_j\} = \{Z_j,X_jX_{j+1},X_1,X_n\}$ which is the basis used to express the control Hamiltonian and compute nested commutators 

\be
[\mathfrak{h}_j,\mathfrak{a}_k]=\sum_{l=1}^d\lambda_{kl}^j\mathfrak{a}_l
\ee

then extend the set $\{\mathfrak{a}_i\}$ so that it is closed under this relation, the resulting set was shown to be \cite{iqc2024}

\be 
Z_i 
\ee 
for $i \in [1,n]$, 
\bqa X_jZ_{jk}X_{j+k+1}\\ Y_jZ_{jk}Y_{j+k+1}\\ X_jZ_{jk}Y_{j+k+1}\\ Y_jZ_{jk}X_{j+k+1}
\eqa 
for $j\in[1,n-k-1]$, 
\bqa
Z_{0j}X_{j+1}\\ Z_{0j}Y_{j+1}\\ X_{n-j}Z_{(n-j)j}\\ Y_{n-j}Z_{(n-j)j}
\eqa with $j\in[0,n-1]$ and finally 
\be \prod_{i=1}^nZ_i\ee where $Z_{jk}=\prod_{i=1}^kZ_{i+j}$.

This turns out to be the same set as the Lie algebra generated by nested commutators of the control Hamiltonian with itself. 

\section{Minimisation of Infidelity}

In this section we will detail the minimisation of the infidelity. In order to have a numerical implementation with polynomial complexity, we represent our operators as coefficient vectors in a space spanned by the set $\{\mathfrak{a}_j\}$. For example the invariant ${\cal I}(T) = \sum_ja_j(T)\hat{\mathfrak{a}}_j$ only requires a polynomial size vector $\mathbf{a}(t)$ and the equation of motion can be written as in equation \ref{eq:deq}:

\be
\dot{\mathbf{a}}(t) = -iK(t)\mathbf{a}(t)
\ee

where $K$ is the adjoint action of the Hamiltonian $K\mathbf{a} = [H,{\cal I}]$. The infidelity can be written as:

\be
{\cal J}({\cal I},{\cal I}_T)=1-\frac{\mbox{Tr}({\cal I}{\cal I}_T)}{\mbox{Tr}({\cal I}_T^2)} = 1 - \frac{\mathbf{a}_T\cdot\mathbf{a}(T)}{\lVert\mathbf{a}_T\rVert^2}\ ,
\ee

where $\mathbf{a}_T$ is the target. The solution for the invariant is given by ${\cal I}(t) = U(t){\cal I}U^\dagger(t)$ and for our specific implementation we use piece-wise constant control pulses and split up the propagator into a product of propagators such that each propagator corresponds to an interval in which the pulse is constant 
 
\be
{\cal I}(t) = U(t_m,t_{m-1})\ldots U(t_1,t_0){\cal I}U^\dagger(t_1,t_0)\ldots U^\dagger(t_m,t_{m-1}).
\ee

Since the Hamiltonian is constant in each interval, each propagator can be written as $U_j = e^{-iH_jt}$. If the invariant is expanded over the set $\{\mathfrak{a}_j\}$, ${\cal I}(t) = \sum_ja_j(t)\hat{\mathfrak{a}}_j$ then the solution in terms of the vector of time dependent coefficients can be written as 

\be
\mathbf{a}(t) = \prod_{l=M}^1e^{-iK_l\Delta t}\mathbf{a}(0),
\ee

where $K_l$ is the adjoint action of the Hamiltonian in the time interval $l$, $K_l\mathbf{a} = [H_l,{\cal I}]$ which is also defined in equation \ref{eq:deq}. The operator $K_l$ can be numerically constructed via the structure constants, as given by equation \ref{eq:deq}, so no reference to the full Hilbert space is needed. The gradient of ${\cal I}(t)$ can be approximated as 

\begin{align}
\frac{\partial\dot{\mathbf{a}}(t)}{\partial f^q_p} &= \left(\prod_{l=m}^{q-1}e^{-iK_l\Delta t}\right)\frac{\partial e^{-iK_q\Delta t}}{\partial f_p^q}\left(\prod_{l=q+1}^{1}e^{-iK_l\Delta t}\right)\mathbf{a}(0) \\
\frac{\partial e^{-iK_q\Delta t}}{\partial f_p^q} &\approx \left(\Delta tK_p - \frac{\Delta t^2}{2}[K_q,K_p] \right)e^{-iK_q\Delta t}
\end{align}

where $K_q$ is the adjoint form of the Hamiltonian in time bin $q$ and $K_p$ is the part of $K_q$ corresponding to the driving $f_p^q$. Using this explicit dependence of the infidelity on the control pulses, the MATLAB function fmincon can perform the minimisation. $10n$ time bins are used for the pulses and an initial guess for the duration of $\frac{n\pi}{2}$, and $\Delta t$ is given by the duration divided by the number of bins. We also include the duration as a optimisation parameter with a gradient computed by finite differences.

\begin{figure}
    \centering
    \includegraphics[width=\linewidth]{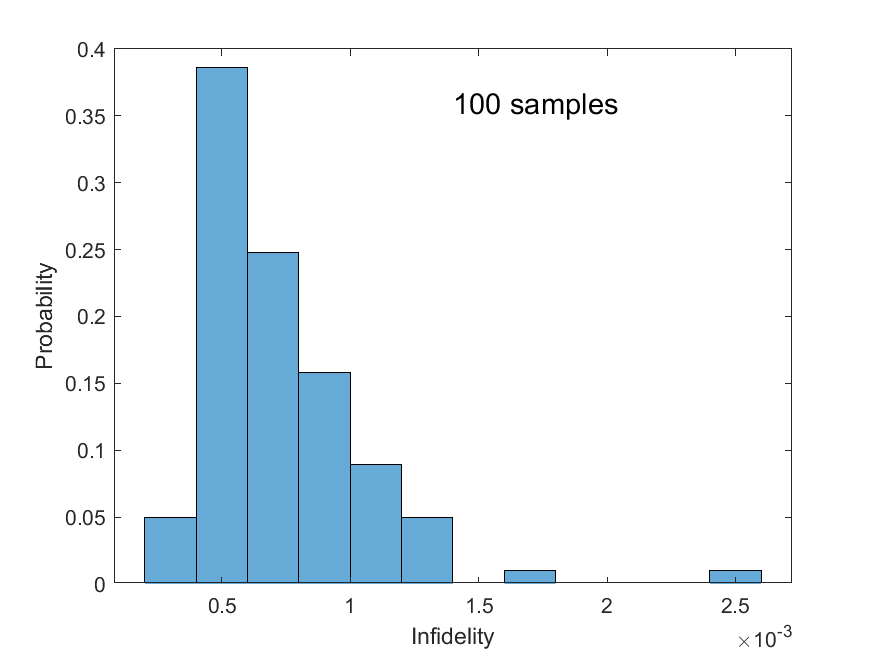}
    \includegraphics[width=\linewidth]{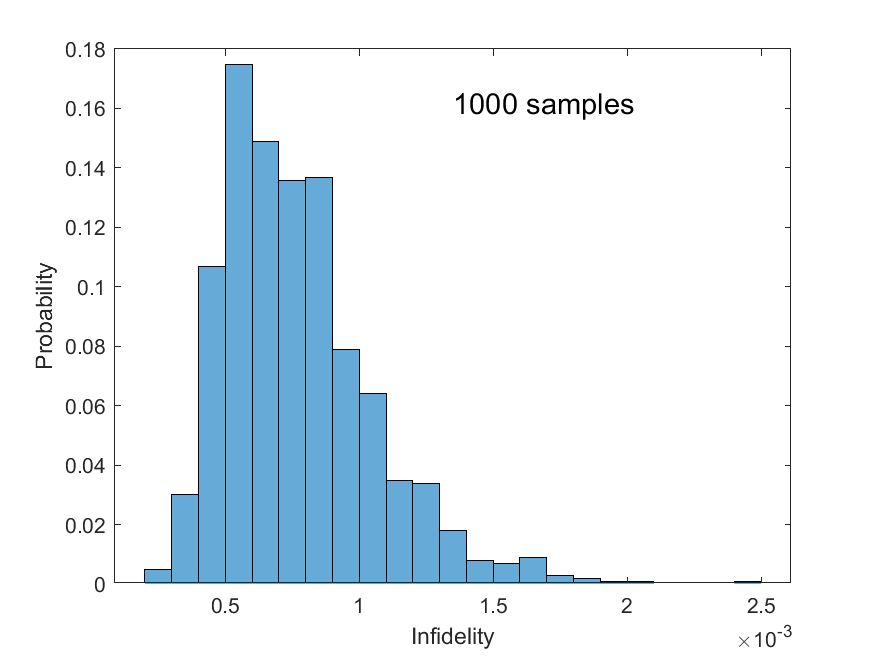}
    \caption{Infidelity distribution is consistent for averages over 100 and 1000 samples.}
    \label{fig:100samples}
\end{figure}

For robustness the same technique is used except the fidelity is averaged over an ensemble of invariants ${\cal I}_i$ where each has a slightly different trajectory and final solution ${\cal I}_i(T)$ due to the small deviations in $H_i$. For every 50 iterations in the optimisation, a new set of errors is chosen so that overfitting to a specific sample can be avoided, and the control pulses achieve robustness against any error within the error hypercube $[5\%,-5\%]^{n-1}$. Every 50 iterations the average infidelity is verified with a sample of 100 points, figure \ref{fig:100samples} shows the typical distribution of fidelities for 100 and 1000 points to demonstrate that 100 is a large enough sample to measure the mean.

\section{Minimisation of Perturbative Correction}

To minimise the perturbative correction we minimise the sum ${\cal J} + \mathrm{w}\ {\cal C}_{robust}$ where ${\cal J}$ is defined as the nonrobust infidelity used in section \ref{sec:rob}, ${\cal C}_{robust}$ is the correction for the solution that is subject to optimisation, and $\mathrm{w} = ({\cal C}_{nonrobust})^{-1}$ is a weight calculated from the correction (\ref{eq:constraint}) for a nonrobust solution to the problem. For the $ZZ$ error, the explicit perturbative correction is 

\begin{align}
{\cal I}_1(t) &= {\cal I}_1(0) -i\int_0^t[\tilde H_P(t_1),{\cal I}_1(0)]dt_1\\ &= {\cal I}_1(0) -i\sum_j\lambda_j\int_0^t[U^\dagger Z_jUU^\dagger Z_{j+1}U,{\cal I}_1(0)]dt_1\\ &= {\cal I}_1(0) -i\sum_{jkl}\lambda_j[\mathfrak{a}_k\mathfrak{a}_l,{\cal I}_1(0)]\int_0^tz_{jk}(t_1)z_{j+1,l}(t_1)dt_1 \\ &= {\cal I}_1(0) -i\sum_{ji}\lambda_j\mathfrak{o}_i\sum_{kl}\eta_{kl}^i\int_0^tz_{jk}(t_1)z_{j+1,l}(t_1)dt_1
\end{align}

Here the coefficient vector $U^\dagger Z_jU = \sum_kz_{jk}(t)\mathfrak{a}_k$ can be written as an expansion over the polynomially scaling set $\{\mathfrak{a}_k\}$ and the commutator $[\mathfrak{a}_k\mathfrak{a}_l,{\cal I}_1(0)] = \sum_i\eta_{kl}^i\mathfrak{o}_i$ has also been written as an expansion over another set $\{\mathfrak{o}_i\}$ which is not necessarily the same as $\{\mathfrak{a}_k\}$. The $\eta_{kl}^i$ is not a structure constant although it looks similar, it only needs to be computed once at the start of the optimisation via parallelised symbolic software. It represents the relative phases of every commutator $[\mathfrak{a}_k\mathfrak{a}_l,{\cal I}_1(0)]$, since many of them give the same operator but with distinct phases. A separate mathematica code is used for computation of $\eta_{kl}^i$. This is done by enumerating every permutation of terms in the commutator $[\mathfrak{a}_k\mathfrak{a}_l,{\cal I}_1(0)] = \sum_i\eta_{kl}^i\mathfrak{o}_i$ and an array is produced so that the numerical minimisation code can quickly read $\eta_{kl}^i$ without having to compute each commutator. The objective function does not include the operators $\mathfrak{o}_i$ but rather groups and sums up all non-zero coefficients of those operators:

\be
{\cal C} = \sum_{ji}\left|\sum_{kl}\eta_{kl}^i\int_0^T z_{jk}(t')z_{j+1,l}(t')dt'\right|^2.
\ee

Grouping them up like this instead of minimising each integral $\int_0^tz_{jk}(t_1)z_{j+1,l}(t_1)dt_1$ makes the approach much less restrictive for the control pulses in $H$. The coefficient vector $z_{jk}(t)$ is constructed similar to the previous work \cite{iqc2024} where piece-wise constant driving is used so $H$ is constant in each interval and the vector is expressed with a given interval as:
\be
z_j(t_n) = \prod_{l=1}^ne^{iK_l\Delta t}z_j(0),
\ee
where $K_l$ is the adjoint action of the Hamiltonian $H_l$ in the time interval $l$ and $z$ is again the vector of coefficients of $U^\dagger Z_jU$ expanded over $\{\mathfrak{a}_j\}$ just as in equation \ref{eq:expand}. This solution is exact because the Hamiltonian is constant within each interval, and the total propagator is split up as a product of propagators where each corresponds to a time bin (40 bins) and $\Delta t$ is just the total duration divided by the number of bins. The total duration is initialised as $\frac{n\pi}{2}$ and the respective gradient is calculated via finite differences so the duration can change during the optimisation. The gradient $\frac{\partial z_{jk}(t)}{\partial f^q_p}$ with respect to driving $p$ in time bin $q$ can be calculated analytically but owing to the high dimensionality it is more efficient to use finite differences over a large number of parallel processors.

\vspace{1cm}

%

\end{document}